\documentclass[nopubldata]{lpor2014}
\usepackage{gensymb}

\begin{document}
	%
	\titlefigure{}
	
	
	
	\abstract{Two vertical-external-cavity surface-emitting laser
(VECSEL) systems producing ultraviolet (UV) radiation at 235\,nm and 313\,nm are demonstrated. The systems are suitable for quantum information processing applications with trapped
beryllium ions. Each system consists of a compact, single-frequency, continuous-wave VECSEL producing
high-power near-infrared light, tunable over tens of nanometers. One system generates 2.4\,W at 940\,nm, using a gain mirror based on GaInAs/GaAs quantum wells, which is converted to 54\,mW of 235\,nm light for photoionization of neutral beryllium atoms. The other system uses a novel gain mirror based on GaInNAs/GaAs quantum-wells, enabling wavelength extension with manageable strain in the GaAs lattice.  This system generates 1.6\,W at 1252\,nm, which is converted to 41\,mW of 313\,nm light that is used to laser cool trapped $^{9}$Be$^{+}$ ions and to implement quantum state
preparation and detection.  The 313\,nm system is also suitable for implementing high-fidelity quantum gates, and more broadly, our results extend the capabilities of VECSEL systems for applications in atomic, molecular, and optical physics.
	}
	
	\title{VECSEL systems for quantum information processing with trapped beryllium ions}
	%
	\titlerunning{}
	\author{S. C. Burd\inst{1,2}, J.-P. Penttinen\inst{3,4}, P.-Y. Hou\inst{1,2}, H. M. Knaack\inst{1,2}, S. Ranta\inst{3,4}, M. M{\"a}ki \inst{3,4}, E. Kantola\inst{3,4}, M. Guina\inst{3,4}, D. H. Slichter\inst{1}, D. Leibfried\inst{1}, and A. C. Wilson\inst{1,*}}

	\authorrunning{S.C. Burd et al.}
	%

    \institute{%
    Time and Frequency Division, National Institute of Standards and Technology, Boulder, Colorado 80305, USA
    \and
    Department of Physics, University of Colorado, Boulder, CO, 80309, USA.
    \and
    Optoelectronics Research Centre, Tampere University, 33720 Tampere, Finland
    \and
    Vexlum Ltd, Korkeakoulunkatu 3, 33720 Tampere, Finland}
	
	\mail{\email{andrew.wilson@nist.gov}}
	%
	\keywords{}
	%
	\maketitle
	\section{Introduction}
	
	Quantum information processing (QIP) technology based on atomic physics is steadily emerging from research laboratories and moving into commercial development \cite{bruzewicz2019}. Often the selection of the atomic system for a particular QIP application is determined more by the availability of suitable lasers than by atomic properties. This is especially apparent with atomic species requiring ultraviolet (UV) laser sources, whose cost and reliability present obstacles. One important example is the beryllium ion ($^{9}$Be$^{+}$), which has several properties advantageous for QIP. The low ion mass helps to achieve high secular trapping frequencies \cite{Wineland1998}, allowing for faster quantum gates and ion transport operations\cite{Bowler2012,Walther2012}, along with stronger Coulomb-mediated coupling between ions in separated-trap arrays\cite{wilson2014tunable,brown2011coupled}. Comparable high-fidelity two-quantum bit (qubit) gates can typically be implemented with less laser intensity than is needed for heavier ion species \cite{Ozeri2007}. Reducing laser requirements may be critical for scaling to larger processors \cite{Wineland1998,Steane2004,Kielpinski2002,Monroe2013,mehta2016integrated} and eventual fault-tolerant operation. Another attractive feature of $^{9}$Be$^{+}$ ions is the $\sim$1.3\,GHz ground state hyperfine splitting, accessible with low-cost microwave electronics and relatively simple antennas. Qubits stored in $^{9}$Be$^{+}$ hyperfine states have exhibited coherence times of several seconds \cite{Langer05} and single-qubit gate errors of $2.0(2)\times10^{-5}$ \cite{Brown2011}. Furthermore, two-qubit gates between $^{9}$Be$^{+}$ hyperfine qubits have been demonstrated with an error of $8(4)\times10^{-4}$ \cite{Gaebler2016}, the lowest reported in any physical system to date.

	\begin{figure}
		\centering
		\includegraphics[width=9cm]{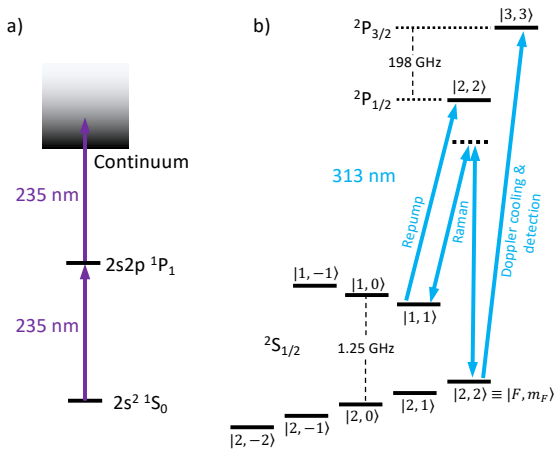}
		\caption{Energy levels relevant to QIP with beryllium ions. a) Energy levels of neutral $^{9}$Be for photoionization with 235\,nm light. b) Energy levels of $^{9}$Be$^{+}$. Light near 313\,nm is used for laser cooling, state preparation and detection, repumping, and multi-qubit quantum gates based on stimulated Raman transitions. Additional repump light resonant with $^{2}S_{1/2} |2,1\rangle \, \leftrightarrow\, ^{2}P_{1/2} |2,2\rangle$ is omitted for clarity. }\label{energy_levels_acw_pyh}
	\end{figure}
    
    Beryllium QIP experiments typically use laser light at 235\,nm to photoionize neutral atoms, and at 313\,nm for laser cooling, state preparation, quantum gates, and measurement. The relevant features of the $^{9}$Be$^{+}$ energy level structure are shown in Figure \ref{energy_levels_acw_pyh}. Light at 235\,nm is often generated by nonlinear frequency conversion of lasers operating at 940\,nm, including titanium-sapphire lasers (both pulsed and CW), and semiconductor lasers \cite{Hsiang2014}. Light at 313\,nm has been generated with nonlinear frequency conversion of light from dye lasers \cite{Brewer1988}, and more recently from fiber lasers\cite{Wilson2011,Hsiang2014} and semiconductor diode lasers \cite{ball2013high,cozijn2013laser,carollo2017third}. Very recently, laser control of beryllium ions was demonstrated using a spectrally-tailored optical frequency comb near 313\,nm \cite{Paschke2019}.
	
Vertical-external-cavity surface-emitting lasers (VECSELs), also known as optically-pumped semiconductor disk lasers offer are a promising alternative to the approaches listed above \cite{Kuznetsov1997}. VECSELs combine the advantages of solid-state external-cavity disk lasers with those of quantum-well semiconductor lasers, and have made considerable progress over the last decade \cite{guina2017optically}. Due to the high quality of the external cavity, photons of the lasing mode have lifetimes that are much longer than the upper-state lifetime of the gain medium. Furthermore, amplified spontaneous emission is suppressed by the cavity and therefore the broad spectral pedestal due to amplified spontaneous emission that is generally present in fiber lasers, diode lasers, and tapered-amplifier systems is largely absent in VECSELs \cite{myara2013}. The external cavity geometry of VECSELs enables both high-power and single-frequency operation, with near-diffraction-limited output beam quality, in a relatively compact package. The semiconductor gain material permits a wide tuning range and can be engineered for emission over a broad range of wavelengths. Additionally, the laser used to optically pump the gain material is not constrained to a narrow absorption band as is typical in solid-state lasers. The relatively high power and beam quality of VECSELs emitting in the near-infrared can allow efficient nonlinear frequency conversion to the visible and the UV.  Previously, we demonstrated VECSEL systems for the generation and manipulation of trapped magnesium ions \cite{Burd:16}. Here, we extend this approach by demonstrating VECSELs for the generation and manipulation of trapped beryllium ions, presenting two new VECSEL-based laser systems that generate 235\,nm light and 313\,nm light.

	\section{Laser systems and characterization}
	
	 \begin{figure}[h]
	\centering\includegraphics[width=9cm]{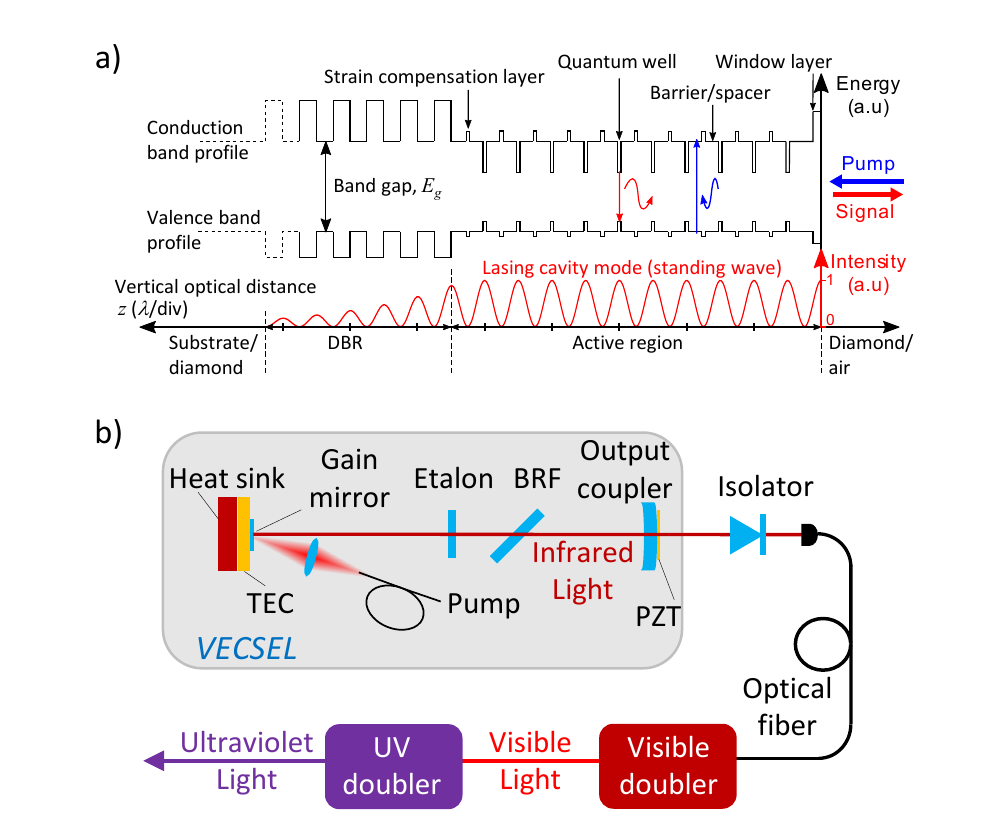}
	\caption{Diagram of the VECSEL-based ultraviolet laser system. a) Schematic semiconductor band gap energy structure of a VECSEL gain mirror. b) VECSEL cavity configuration with two stages of external frequency doubling.}
	\label{Fig:VECSEL_structure}
	\end{figure}
	
	\subsection{Semiconductor gain mirror growth and packaging}
	
	Each VECSEL contains a gain mirror composed of a distributed Bragg reflector (DBR) mirror and an active region with strain compensation layers, quantum wells (QWs), barrier/spacer layers, and a window layer, as shown in the schematic band diagram in Figure \ref{Fig:VECSEL_structure} a. The pump photons (shown in blue, having energy higher than the band gap $E_g$ of the barriers) are absorbed in the structure and generate charge carrier pairs in the valence and conduction bands. These carriers are captured in the quantum wells via diffusion and fast thermal relaxation, and then recombine emitting lower energy signal photons (shown in red). The window layer, having a larger band gap than the barriers, prevents non-radiative recombination in the semiconductor surface, whereas the tensile strain compensation layers reduce the cumulative compressive strain of the quantum wells. The DBR is designed for the laser signal wavelength and locks the phase of this optical field inside the gain mirror (intensity plotted with red below the band gap diagram). As the gain of a quantum well is proportional to the optical field intensity, the quantum wells are placed in the intensity maxima of the lasing cavity mode (i.e. in the anti-nodes of the resonator standing wave). This ordering is called resonant periodic gain (RPG) and maximizes the total gain of the gain mirror. A more detailed description of VECSEL gain mirror technology and material systems can be found for example in \cite{tropper2006,guina2017optically}. Both gain mirrors used in this work, for the 940\,nm VECSEL in the 235\,nm system and for the 1252\,nm VECSEL in the 313\,nm system, were grown using solid-source molecular beam epitaxy (MBE) and are based on GaIn(N)As QWs embedded in GaAs barriers. With this material system we can construct lasers with wavelengths from the low-900\,nm to high-1200\,nm range with more than 1\,W of output power.
	 
	 Due to the substantial wavelength difference between the two VECSEL systems described here, the gain mirror designs also have different layer structures. The active region of the 940\,nm device is comprised of 24 GaInAs QWs distributed in the anti-nodes of the calculated resonator standing wave pattern inside the gain mirror, with two QWs per anti-node (12$\times$2). The total number of QWs is relatively large in this case due to the shallow, low indium-content, GaInAs QWs in the GaAs barriers. The reduced carrier confinement and the resulting poor efficiency is the main mechanism limiting the minimum wavelength of the GaInAs/GaAs material system \cite{chilla2007}. 
	 
	 In comparison, the 1252\,nm gain mirror contains only 10 GaInNAs (so-called dilute-nitride) QWs, again distributed in the anti-nodes of the calculated resonator standing wave pattern, two per anti-node (5$\times$2). The high indium-content QWs at 1252\,nm have good confinement for the carriers so that fewer QWs are needed than at 940\,nm. The number of anti-node regions, 12 in the 940\,nm gain mirror compared to 5 in the 1252\,nm one (number ratio of 2.4), is however more dependent on the pump absorption length, and the physical thicknesses of the two different active regions are similar (ratio of 1.4) due to the wavelength dependence. Without nitrogen, the high indium content at 1252\,nm would deteriorate the crystal quality due to the high cumulative strain caused by the large lattice mismatch between the QWs and the GaAs barriers. However, insertion of a small percentage of nitrogen into the QWs enables the growth of these structures by directly decreasing the lattice mismatch, and also by decreasing the band-gap energy, so that a smaller concentration of indium can be used. As poor crystal quality results in poor optical properties, the cumulative strain is the main limiting mechanism for the upper wavelength of the GaInAs/GaAs system \cite{guina2017optically}. Dilute nitrides extend this range, but growing them is technically challenging \cite{korpijarvi2015high}. Consequently, as described below, the 1252\,nm gain mirror delivers nearly twice the single-frequency output power at 1231\,nm than at the target wavelength of 1252\,nm, suggesting room for future improvement.
	 
	For proper VECSEL operation, the temperature of the gain mirror must be stabilized, and the heat from pump laser dissipation must be removed. With the 940\,nm gain mirror we used a standard flip-chip cooling method where the heat flows from the active region through the DBR to the heat spreader. The gain mirror was grown active-region-first on a GaAs substrate, followed by the GaAs/AlAs DBR. The gain mirror was then diced into 2.5$\times$2.5\,mm$^{2}$ chips and the DBR back surface of each chip was bonded to a 3$\times$3$\times$0.3\,mm$^{3}$ synthetic multi-crystal diamond heat-spreader. After bonding, the GaAs substrate was removed with a combination of mechanical lapping and wet etching. An ion-beam-sputterred (IBS) anti-reflection (AR) coating was applied to the top surface of the approximately 5\,$\mu$m-thick semiconductor structure for reduced reflection of pump and laser light and for protection of the gain chip surface. Finally, the diamond back surface was soldered to a gold-plated copper heat sink for efficient heat extraction.
	
	For the 1252\,nm gain mirror, we chose a different cooling geometry  better suited to devices operating at longer wavelengths that are challenging to grow. The 1252\,nm gain mirror (with active and DBR regions) was first diced into 3$\times$3\,mm$^{2}$ chips and the active-region surface of each was liquid-capillary bonded to a transparent heat spreader. The advantage of this so-called intra-cavity cooling configuration is that the low-thermal-conductivity DBR is not in the heat path between the hot active region and the heat spreader. Conversely, since the intra-cavity beam now passes through the heat spreader, this component has to be of good optical quality. We used a 3$\times$3$\times$0.25\,mm$^{3}$ synthetic single-crystal diamond heat spreader, polished with a 2-degree angle between facets to suppress spurious etalons. The top surface of the diamond was soldered to a copper heatsink with an opening for the intra-cavity beam to access the gain mirror. Finally, the diamond top surface was AR-coated (using the IBS method) to minimize intra-cavity loss and etaloning. We obtained better results at 1252 nm with intra-cavity-cooled gain mirrors than with flip-chip-cooled gain mirrors. For the 940\,nm gain mirrors, we tested only flip-chip devices, which are often preferred because multi-crystal diamond heat spreaders are more easily sourced and lower cost than single-crystal ones.
	
    Both VECSEL cavities (Figure \ref{Fig:VECSEL_structure} b) consist of a gain mirror and an output-coupler mirror (OC, Layertec\footnote[1]{\label{footnote:note1}To foster understanding, a range of commercial components and suppliers are identified in this paper. Such identification does not imply recommendation or endorsement by the National Institute of Standards and Technology, nor does it imply that the components identified are necessarily the best available for the purpose.}, $\sim 2\,\%$ transmission, 200\,mm radius of curvature, 12.7\,mm diameter), spaced approximately 125\,mm apart (giving a $\sim$1.2\,GHz free spectral range (FSR)). To provide single-frequency operation and coarse tuning, we used the following intra-cavity elements: a Brewster-angled birefringent filter (BRF, single quartz plate, 3\,mm thickness, New Light Photonics, model BIR0030) and an etalon (yttrium aluminium garnet, 1\,mm thickness, Light Machinery, model OP-3167-1000).  Both these elements are actively temperature-stabilized. For fine tuning and stabilization of the VECSEL output frequency, the small OC mirror is mounted on a ring-shaped piezo-electric transducer (PZT, CTS Ceramics, model NAC2123-A01), which allows the cavity length to be adjusted. The gain mirror is optically pumped by a high-power, multi-mode diode laser (typically, Coherent, model M1F2S22-807.3-40C-SS2.6T3) emitting near 808\,nm, which is fiber coupled and focused (Oz Optics, model LPF-D4-808-200/240-QM-2.2-12-6.2AS,13.9AS-35-5HPL-3A-1.5) to produce a $\sim 200 \, \mu$m beam waist (radius) on the gain-mirror surface. Although the pump absorption band of the gain material is very broad, high-power diode lasers emitting near 808\,nm were selected, as these are readily available and relatively inexpensive, and the gain mirror pump absorption length was optimized accordingly. The gain mirror copper heatsink is mounted on a thermo-electric cooler (TEC) for temperature stabilization and control. The TEC is water-cooled using a micro-channel heat exchanger and a low-vibration chiller (important for narrow-linewidth laser operation). The VECSEL cavity components are mounted on an Invar (low coefficient of thermal expansion) baseplate that is housed in an O-ring-sealed enclosure for stable operation. We have found that removing intra-cavity water vapor, by including a desiccant inside the laser enclosure and purging with dry nitrogen, improves frequency stability and output power at wavelengths where water absorption lines are present.
    
    Pump alignment is performed by imaging the spontaneous emission and scattered pump light from the gain chip using a CMOS camera and a high magnification zoom lens (Thorlabs, model MVL7000) with neutral-density filters to attenuate the light as needed. The distance between the pump focusing optics and the gain chip is adjusted to achieve the targeted spot diameter and intensity profile. Gaussian or super-Gaussian intensity profiles with a laser-mode-to-pump-spot ratio larger than 0.8 are typically preferred for single-mode operation \cite{Laurain2019}. Coarse alignment is performed, with the OC, BRF, and etalon removed, by retro-reflecting a beam from an external visible alignment laser from the gain chip surface. The OC is then inserted and adjusted so that the secondary reflection from the OC is reflected back onto the gain mirror. Finer cavity adjustment to establish lasing is performed by monitoring the output just beyond the OC mirror with a high-gain, large-area photodiode (Thorlabs, model PDA50B2).  Long-pass filters placed before the photodiode strongly attenuate scattered pump light while transmitting spontaneous emission and laser output. Lasing is achieved by adjusting the OC mirror to increase the photodiode signal. The multi-mode laser output power (without BRF and etalon installed) is optimized not only by adjusting the pump beam and OC mirror alignments but also the gain mirror set-point temperature. The Brewster-angled BRF is then inserted into the cavity and the OC mirror is tilted to correct for the BRF's displacement of the intra-cavity beam and to re-establish lasing (using the photodiode method described above). Finally, the etalon is installed and aligned to achieve lasing again.
    
	In single-mode operation of the VECSELs, coarse-range tuning (in $\sim$80 GHz steps over $\sim$10\,THz) is achieved by BRF rotation, as well as by adjusting the BRF set-point temperature (0.1\,THz--1\,THz range at $\sim$37 GHz/K). Intermediate-range tuning is achieved by tilting the etalon and adjusting its temperature (1\,GHz--100\,GHz range at {$\sim$\,-3\,GHz/K}, $\sim$80\,GHz FSR). Fine tuning is achieved by adjusting the laser cavity length using the output-coupler PZT (0.01\,GHz--1\,GHz, $\sim$30\,MHz/V). By simultaneously tuning the etalon temperature and the laser cavity length, it should be possible to achieve mode-hop-free tuning over $\sim$7\, GHz (equivalent to approximately six times the cavity FSR) limited by the $\sim$3 $\mu$m maximum travel of the cavity PZT.
	
	\subsection{The 235\,nm laser source}
\begin{figure}[h]
		\centering\includegraphics[width=8cm]{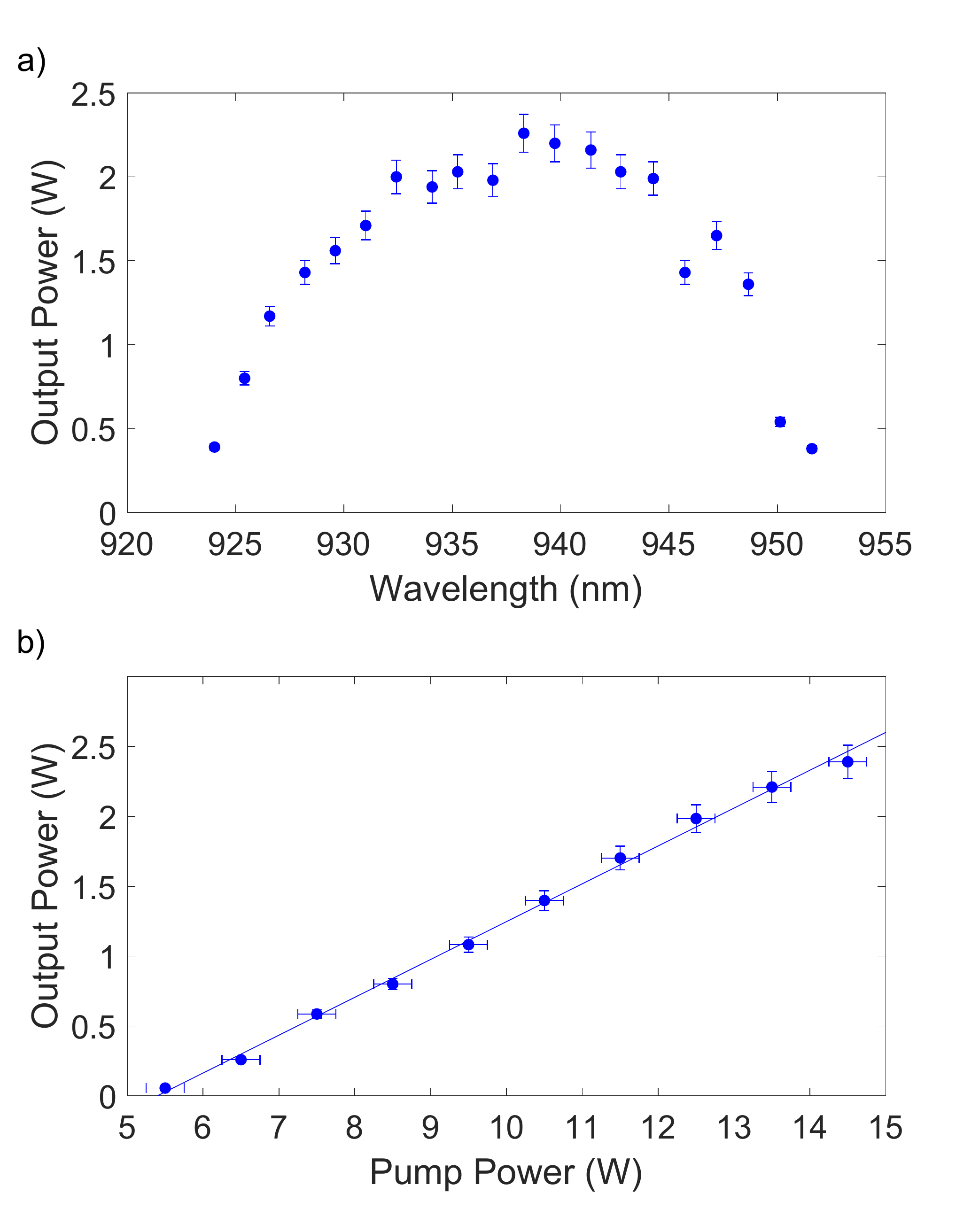}
		\caption{a) Single-mode output power at 940\,nm as a function of wavelength with 14.5\,W pump power, obtained by rotating the BRF and tuning etalon temperature at each wavelength. b) Output power at 940\,nm versus 808\,nm pump power. A linear fit to the data (solid line) gives a slope efficiency of 27(1)\,\%.  Power measurement uncertainty is 5\,\% and the wavelength accuracy uncertainty is 0.002\,nm.}
		\label{fig:Power_940}
	\end{figure}

    Performance characteristics of the 940\,nm VECSEL are shown in Figure \ref{fig:Power_940}. With 14.5 W of pump power, the tuning range is $\sim\,$30\,nm, and the slope efficiency at 940\,nm is 27(1)\% for single-frequency operation. To estimate the VECSEL linewidth, we analyze the beat note signal between the 940\,nm VECSEL output and that of a free-running titanium-sapphire (TiS) laser (M-Squared Lasers, model SolsTiS, nominal linewidth $<$50\,kHz), with the VECSEL frequency-offset locked to the TiS laser \cite{Castrillo2010}. From the spectral width of the beat signal, we determine the linewidth of the VECSEL to be $<$100\,kHz.  This is considerably less than the linewidth of relevant atomic transitions and sufficiently narrow that frequency fluctuations will not be converted to significant amplitude fluctuations by subsequent resonant frequency doubling stages.
   
\begin{figure}[]
		\centering\includegraphics[width=8cm]{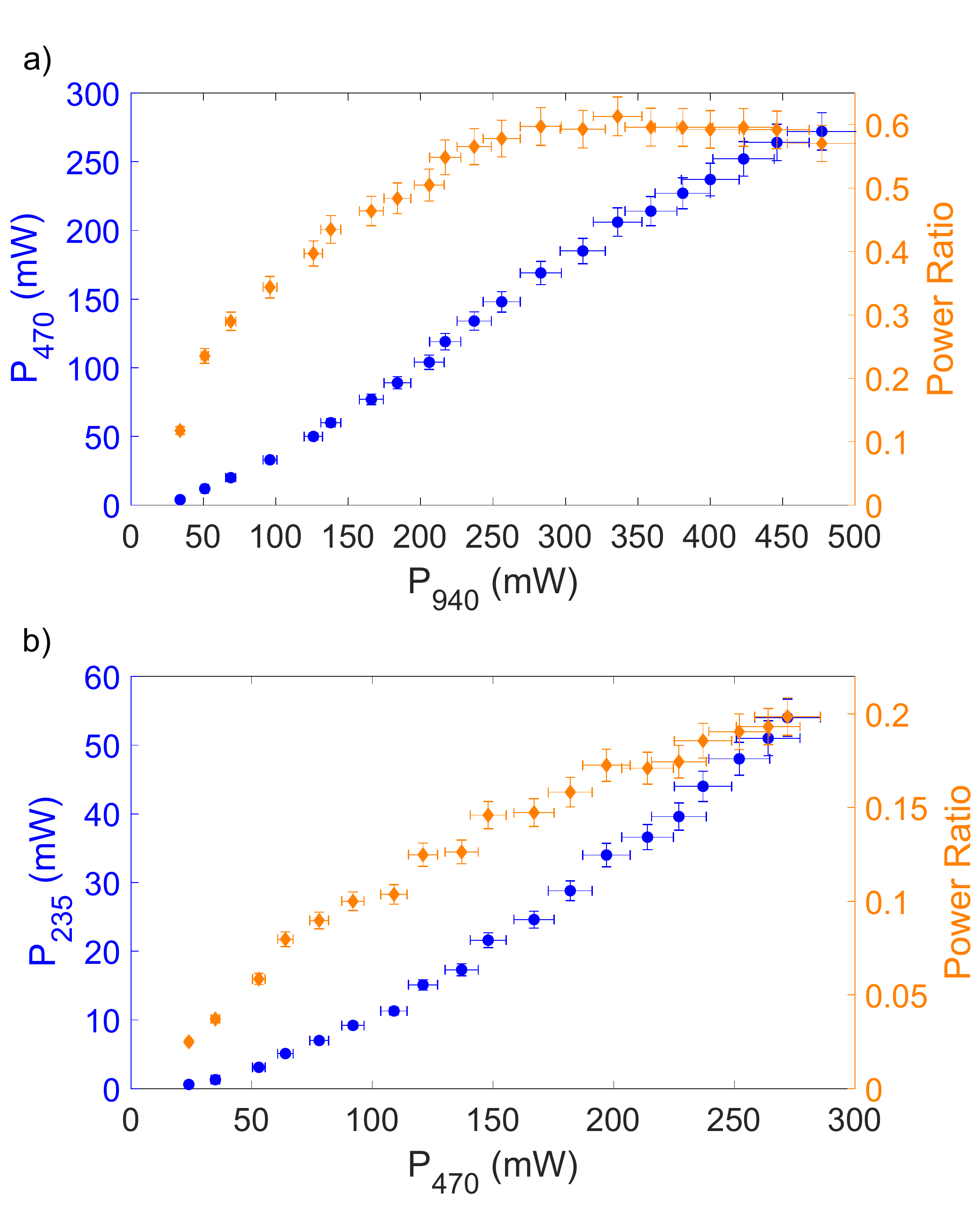}
		\caption{Measured output powers and efficiencies of frequency-doubling stages used in the 940\,nm VECSEL system. a) The 470\,nm power (blue circles), and the 470\,nm to 940\,nm power ratio (orange diamonds), versus the input power at 940\,nm. b) The 235\,nm power (blue circles) and the 235\,nm to 470\,nm  power ratio (orange diamonds) versus the input power at 470\,nm. Uncertainty in all power measurements is  5\,\%.
			}\label{doublers}
	\end{figure}

	The output of the 940\,nm VECSEL is coupled into a high-power polarization-maintaining (PM) optical fiber that delivers light to the first of two second-harmonic generation (SHG) enhancement cavities. This 940\,nm to 470\,nm frequency-doubling cavity is locked to the single-frequency output of the VECSEL using the Pound-Drever-Hall (PDH) method \cite{Drever1983}. Following the design by Hsiang and colleagues \cite{Hsiang2014}, frequency doubling to 470\,nm is implemented using a periodically-poled potassium titanyl phosphate (PPKTP) nonlinear optical crystal (Raicol Crystals, 20\,mm length). The performance of the VECSEL-driven 470\,nm source is shown in Figure \ref{doublers} a. With 0.48(2)\,W at 940\,nm we obtain 0.27(1)\,W at 470\,nm. While this is more than sufficient for our application, we note that the 940\,nm power to the doubling cavity could be further increased. However, cavity-locking instabilities due to apparent thermal lensing in the crystal prevent reliable locking above $\sim$\,0.50\,W at 470\,nm\cite{le200575,Hsiang2014}. 
	
	The second stage of frequency doubling, from 470\,nm to 235\,nm, is implemented using an enhancement cavity design \cite{Hsiang2014}, adapted from an earlier 313\,nm design \cite{Wilson2011}, that uses a Brewster-angled BBO crystal. The performance of the 235\,nm frequency doubling stage is shown in Figure \ref{doublers} b. With 0.27(1)\,W at 470\,nm we obtain 54(3)\,mW at 235\,nm, and a maximum power ratio of approximately 20\,\%. For context, this UV power is approximately 50 times greater than the power we typically use to photo-ionize beryllium atoms to load them into an ion trap.
	
	
	\subsection{The 313\,nm laser source}

\begin{figure}[h]
		\centering\includegraphics[width=8cm]{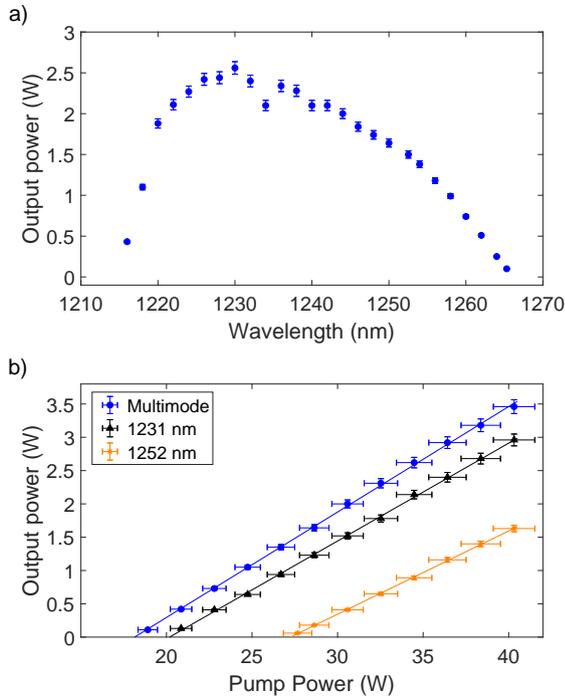}
		\caption{a) Output power of the 1252\,nm VECSEL as a function of wavelength in single-frequency operation  at 40.3\,W pump power, obtained by rotating the BRF and adjusting the etalon and gain chip temperatures at each wavelength to optimize the output power. b) The measured VECSEL output powers, under three different operating conditions, versus the input power of the 808\,nm pump laser. Solid lines are linear fits used to determine slope efficiencies. Uncertainty of power measurement at each data point in a) and b) is 3\,\%.}\label{fig:powercurve}
	\end{figure}

    
    Initial characterization of the 1252\,nm VECSEL was performed by measuring the output power and wavelength tuning range of the laser (Figure \ref{fig:powercurve}). In multi-longitudinal mode operation (with BRF and etalon removed), a maximum output power of 3.5(1)\,W at 1231\,nm, and a slope efficiency of 15.8(3)\,\% was measured (Figure \ref{fig:powercurve} b). In single-frequency operation (with BRF and etalon installed), we measured a reduced slope efficiency of 14.7(3)\,\% and an output up to 3.0(1)\,W at 1231\,nm due to the additional loss from the intra-cavity elements. When tuned to the target wavelength for our application (1252\,nm), a slope efficiency of 12.5(2)\,\% and output up to 1.63(5)\,W were obtained. We are able to tune the VECSEL over a 50\,nm range.  This is limited at lower wavelengths by the available gain chip cooling power. We note that roll-over in the output power at the highest pump power was not observed, suggesting that higher output power could be achieved by increasing the pump power. We anticipate fabricating future gain mirrors with peak gain close to the target wavelength by optimizing the active region thickness and the QW number and composition.
    
    For conversion of 1252\,nm light to 626\,nm, a commercial, fiber-coupled, periodically-poled lithium-niobate waveguide doubler (NTT Electronics Corp, model WH-0626-000-A-B-C) is used. The temperature of the waveguide is controlled by a TEC to maintain the quasi-phase matching condition. For this conversion stage, with 1.63(5)\,W of 1252\,nm power from the VECSEL we obtain a maximum power of 0.53(3)\,W at 626\,nm, and a power ratio of 33(2)\,\% (including loss from the fiber coupling into the waveguide). The maximum power at 626\,nm is limited by the available 1252\,nm power. We note that 2.4\,W at 671\,nm has been generated using a similar waveguide-based SHG device \cite{kretzschmar20172}. 
    
    The 626\,nm light from the waveguide doubler is coupled into a PM optical fiber and delivered to a cavity-enhanced SHG setup \cite{Wilson2011} that uses a Brewster-angled BBO crystal for conversion to 313\,nm. With 0.33(2)\,W input power we obtain 41(2)\,mW at 313\,nm. 
	
	To estimate the spectral linewidth of the 1252\,nm VECSEL, we analyze the beat note signal between the 626\,nm output of the waveguide doubler and the output of a frequency-stabilized fiber-laser-based 626\,nm source similar to that described in \cite{Wilson2011}.  Following the approach described above for the analysis of the 940\,nm VECSEL, the 1252\,nm VECSEL is frequency-offset locked to the fiber-laser source \cite{Castrillo2010}.  From the spectral width of the beat signal, we estimate the linewidth of the 1252\,nm VECSEL to be $<$110\,kHz.  This is also considerably less than the linewidths of relevant atomic transitions and sufficiently narrow that frequency fluctuations will not be converted to significant amplitude fluctuations by the resonant UV frequency doubling stage.

	\section{Tests with beryllium atoms and ions}
	
	The VECSEL systems were tested with neutral $^{9}$Be and $^{9}$Be$^{+}$ ions in two different experimental setups.
	
	Photoionization with the 235\,nm VECSEL system was tested on a thermal beam of neutral beryllium atoms. The thermal beam was generated by resistively heating a length of beryllium wire  spiral-wound onto tungsten support wire. The beam was weakly collimated using a 2.5\,mm diameter aperture. Immediately beyond the aperture, the atomic beam intersected with a perpendicular 235\,nm laser beam near-resonant with the $^{1}S_{0}$ to $^{1}P_{1}$ transition and focused to an intensity of $\sim$80(20)\,kW/m$^2$ at the center of the atomic beam. For comparison, the saturation intensity of this transition is $\sim$8.7\,kW/m$^2$ \cite{Cook2018}. Ions are produced by a two-photon process (see Figure \ref{energy_levels_acw_pyh} a). On resonance, the first photon excites the neutral atom to the $^{1}P_{1}$ state and a second photon excites the electron to the continuum. These ions are counted using a Channeltron electron multiplier (CEM) (Photonis Magnum 5901 Electron Multiplier) with a bias potential of -1.7\,kV.
	
\begin{figure}[h]
		\centering\includegraphics[width=8cm]{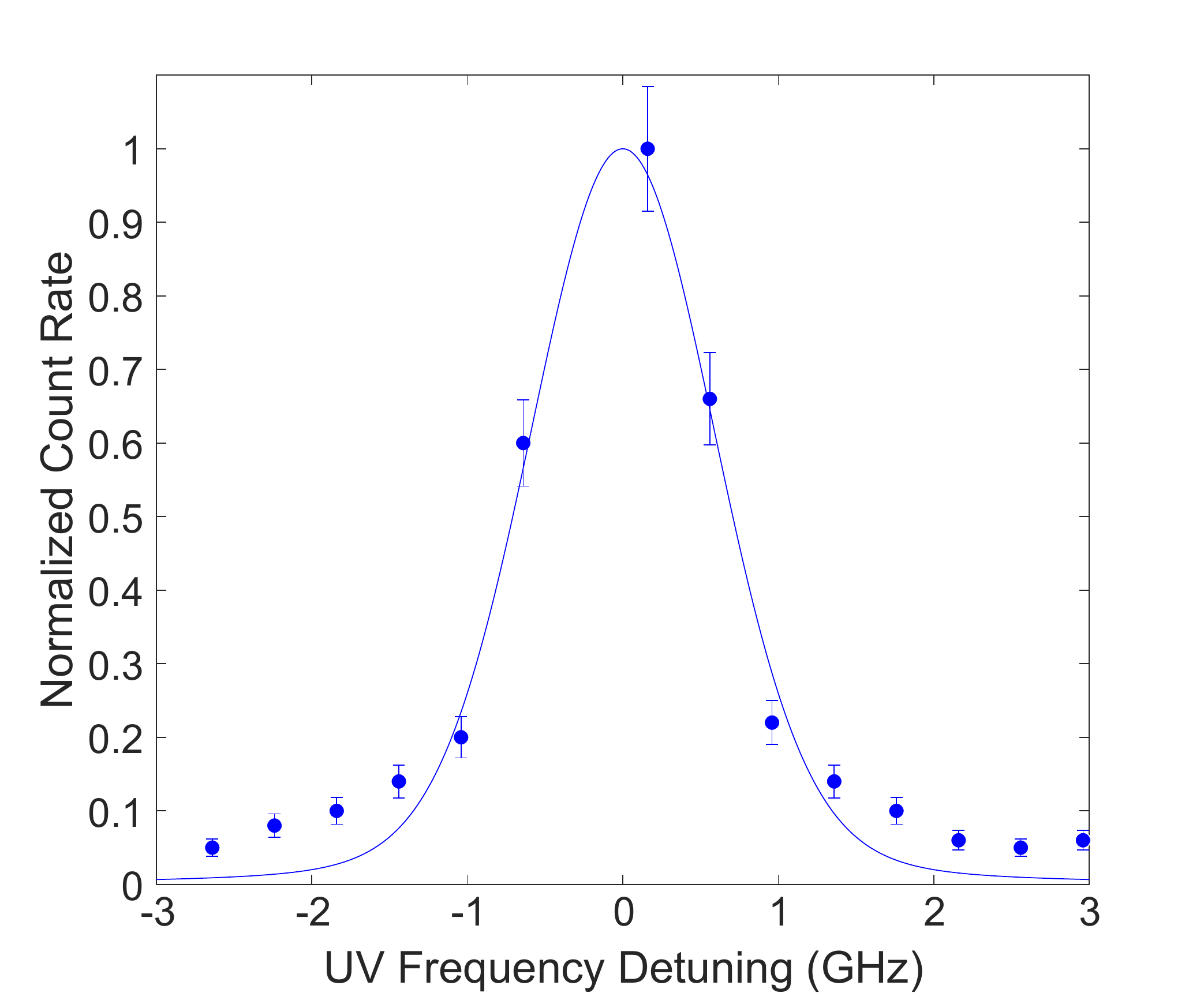}
		\caption{Count rate measured by the CEM as a function of the quadrupled VECSEL (UV) frequency detuning from {1,276.080(2)\,THz.} The central peak is overlaid with a Voigt model.  Count rate measurement uncertainty is estimated to be 4\,\%.}
		\label{Fig:ioncounts}
	\end{figure}

     With the neutral atomic beam flux held constant, we record the ion count rate as a function of the VECSEL frequency to obtain a photoionization lineshape (Figure \ref{Fig:ioncounts}). Maximum photoionization rates are measured at a VECSEL frequency of 319.0200(6)\,THz, corresponding to 1,276.080(2)\,THz in the UV, consistent with the recent precision measurement of the $^{1}S_{0}$ to $^{1}P_{1}$ transition \cite{Cook2018}. The central feature of the photoionization lineshape includes contributions from the natural linewidth of 87(5)\,MHz \cite{Cook2018}, power broadening by a factor of $\sim\,$2.4(3), and  1.3(2)\,GHz of residual Doppler broadening from imperfect collimation of the atomic beam \cite{Demtroder}. A Voigt model \cite{Demtroder} including only these mechanisms shows good agreement with the central feature.  The origin of the weak off-resonant photoionization (the broad pedestal feature of the lineshape) has not been investigated.

\begin{figure}[]
	\centering
	\includegraphics[width=8cm]{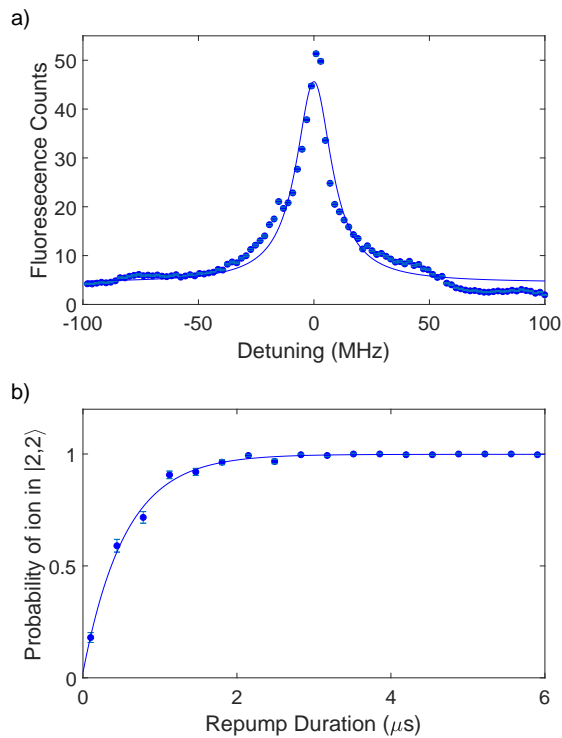}
	\caption{a) Spectroscopy of the  $^{2}S_{1/2}|2,2\rangle \leftrightarrow ^{2}P_{3/2}|3,3\rangle$ cycling transition with UV light from the 313\,nm VECSEL source. Each data point indicates the mean of 300 experimental repetitions each with a 330 $\mu$s detection period. The solid line is a Lorentzian fit to the data. b) Probability of measuring $^{2}S_{1/2}|2,2\rangle$ after repumping from the $^{2}S_{1/2}|1,1\rangle$ state with UV light from the 313\,nm VECSEL source. The solid curve is an exponential fit. Data points indicate the mean of 300 repetitions. Error bars in a) and b) indicate the standard deviation of the mean.}\label{test1252}
	\end{figure}

The 313\,nm laser source was used to perform Doppler cooling, resonant detection and repumping in an established trapped-ion system described in \cite{Tan2014}. For convenience, we stabilized the 626\,nm light produced by the frequency-doubled VECSEL to 626\,nm light produced by a fiber laser system that is part of the trapped-ion setup \cite{Wilson2011}. We used the 313 nm laser source for Doppler cooling single $^{9}$Be$^{+}$ ions loaded in the trap. The frequency-quadrupled light at 313\,nm was then tuned across the $^{2}S_{1/2} |2,2\rangle\leftrightarrow{^{2}P_{3/2}} |3,3\rangle$ cycling transition using an acousto-optic modulator (AOM). Fluorescence photons were collected by an optical imaging system and counted using a photo-multiplier tube (PMT). The resulting spectrum is shown in Figure \ref{test1252} a.  Within laser-intensity uncertainties, the lineshape obtained with the VECSEL source is identical to that obtained with our fiber laser source.

To test repumping with the VECSEL source, the laser is tuned into resonance with the $^{2}S_{1/2}\,\leftrightarrow \,^{2}P_{1/2}$ transition of $^{9}$Be$^{+}$ (Figure \ref{energy_levels_acw_pyh} b). We swept the frequency across the atomic transition using an AOM.  In this experiment, a single $^{9}$Be$^{+}$ ion was prepared in the $|1,1\rangle$ state by first optical pumping to the $|2,2\rangle$ state and then applying a microwave $\pi$ pulse resonant with the $|2,2\rangle\leftrightarrow|1,1\rangle$ transition. Repumping light is then applied for various durations, ideally repumping the ion to the $|2,2\rangle$ state for subsequent fluorescence detection with light resonant with the $^{2}S_{1/2} |2,2\rangle \, \leftrightarrow\, ^{2}P_{3/2} |3,3\rangle$ cycling transition (see Figure \ref{energy_levels_acw_pyh} b). The detection in each experiment yields a binary outcome; the ion either fluoresces significantly or it does not. By setting a threshold of 10 counts, we can reliably assign a 0 or 1 to the outcome of a given experiment. For each duration, the population of
	$|2,2\rangle$ is determined by calculating the mean of the results from 300 experimental repetitions. 
   The repump time {$\tau = 0.55(4)$\,$\mu$s} is determined by fitting the exponential function $ 1-e^{-t/\tau}$, where $t$ is the repumping pulse duration, to the data in Fig. \ref{test1252} b. Accounting for laser-intensity differences, the repumping obtained using the VECSEL system is indistinguishable from that obtained with our fiber laser source.
	
	\section{Conclusion}
	In summary, we have presented two widely-tunable VECSEL-based laser sources capable of implementing tasks required for quantum information processing experiments involving trapped $^{9}$Be$^{+}$ ions. The first system generates up to 2.4\,W single-frequency light at 940\,nm and is frequency doubled twice to generate 235\,nm light for photoionization of neutral Be. The second system produces up to 1.6\,W at 1252\,nm which is converted to 313\,nm by two stages of frequency doubling, and can be used for Doppler cooling, detection, and repumping of trapped $^{9}$Be$^{+}$ ions. We note that the large tuning range of the 940\,nm VECSEL system presented in this work could allow for generation of light near 313\,nm by third harmonic generation \cite{carollo2017third}. The inherent power scalability of the VECSEL design should allow the generation of higher output power \cite{Kuznetsov1997}, which is desirable for mitigating spontaneous emission errors associated with quantum-logic gates using far-detuned stimulated Raman transitions\cite{Ozeri2007}. Further improvements to overall system efficiency could be achieved using intra-cavity second harmonic generation \cite{kantola2014}. Together with previous demonstrations of VECSEL systems for trapped magnesium ions\cite{Burd:16} and also considering the wide range of infrared frequencies accessible to these systems, our work shows that VECSELs, when combined with well-established and efficient frequency conversion techniques, will facilitate quantum information processing technologies based on atomic species requiring ultraviolet laser sources.
	
	\begin{acknowledgement}
	S.C.B, J.-P.P,. P.-Y.H. and H.M.K. all contributed equally to this work. This work was supported by the NIST Quantum Information Program and the Academy of Finland project QUBIT (278388). We thank Stephen Erickson, Jenny Wu, Daniel Cole, and Yong Wan for assistance in the laboratory, and Richard Fox and Jenny Wu for comments on the manuscript. P.-Y.H., H.M.K., and S.C.B. acknowledge support of the Professional Research Experience Program (PREP), operated jointly by NIST and the University of Colorado. J.-P.P acknowledges support of the Jenny and Antti Wihuri Foundation and the Walter Ahlstr{\"o}m Foundation.	
	\end{acknowledgement}
	
	%
	

	\bibliographystyle{lpr}
	\bibliography{ref}

\providecommand{\WileyBibTextsc}{}
\let\textsc\WileyBibTextsc
\providecommand{\othercit}{}
\providecommand{\jr}[1]{#1}
\providecommand{\etal}{~et~al.}


\begin{thebibliography}{[10]}

\bibitem{bruzewicz2019}
 \textsc{C.\,D. Bruzewicz},  \textsc{J.~Chiaverini},  \textsc{R.~McConnell},
  and  \textsc{J.\,M. Sage},
 \jr{Appl. Phys. Rev.} \textbf{6}, 021314 (2019).


\bibitem{Wineland1998}
 \textsc{D.\,J. Wineland},  \textsc{C.~Monroe},  \textsc{W.\,M. Itano},
  \textsc{D.~Leibfried},  \textsc{B.\,E. King},  and  \textsc{D.\,M. Meekhof},
 \jr{J. Res. Nat. Inst. Stand. Technol.} \textbf{103}, 259 (1998).


\bibitem{Bowler2012}
 \textsc{R.~Bowler},  \textsc{J.~Gaebler},  \textsc{Y.~Lin},  \textsc{T.\,R.
  Tan},  \textsc{D.~Hanneke},  \textsc{J.\,D. Jost},  \textsc{J.\,P. Home},
  \textsc{D.~Leibfried},  and  \textsc{D.\,J. Wineland},
 \jr{Phys. Rev. Lett.} \textbf{109}, 080502 (2012).


\bibitem{Walther2012}
 \textsc{A.~Walther},  \textsc{F.~Ziesel},  \textsc{T.~Ruster},  \textsc{S.\,T.
  Dawkins},  \textsc{K.~Ott},  \textsc{M.~Hettrich},  \textsc{K.~Singer},
  \textsc{F.~Schmidt-Kaler},  and  \textsc{U.~Poschinger},
 \jr{Phys. Rev. Lett.} \textbf{109}, 080501 (2012).


\bibitem{wilson2014tunable}
 \textsc{A.\,C. Wilson},  \textsc{Y.~Colombe},  \textsc{K.\,R. Brown},
  \textsc{E.~Knill},  \textsc{D.~Leibfried},  and  \textsc{D.\,J. Wineland},
 \jr{Nature} \textbf{512}, 57 (2014).


\bibitem{brown2011coupled}
 \textsc{K.\,R. Brown},  \textsc{C.~Ospelkaus},  \textsc{Y.~Colombe},
  \textsc{A.\,C. Wilson},  \textsc{D.~Leibfried},  and  \textsc{D.\,J.
  Wineland},
 \jr{Nature} \textbf{471}(7337), 196 (2011).


\bibitem{Ozeri2007}
 \textsc{R.~Ozeri},  \textsc{W.\,M. Itano},  \textsc{R.\,B. Blakestad},
  \textsc{J.~Britton},  \textsc{J.~Chiaverini},  \textsc{J.\,D. Jost},
  \textsc{C.~Langer},  \textsc{D.~Leibfried},  \textsc{R.~Reichle},
  \textsc{S.~Seidelin},  \textsc{J.\,H. Wesenberg},  and  \textsc{D.\,J.
  Wineland},
 \jr{Phys. Rev. A} \textbf{75}, 042329 (2007).


\bibitem{Steane2004}
 \textsc{A.\,M. Steane},
 \jr{Quantum Inf. Comput.} \textbf{7}, 171 (2004).


\bibitem{Kielpinski2002}
 \textsc{D.~Kielpinski},  \textsc{C.~Monroe},  and  \textsc{D.\,J. Wineland},
 \jr{Nature} \textbf{417}, 709 (2002).


\bibitem{Monroe2013}
 \textsc{C.~Monroe} and  \textsc{J.~Kim},
 \jr{Science} \textbf{339}, 1164 (2013).


\bibitem{mehta2016integrated}
 \textsc{K.\,K. Mehta},  \textsc{C.\,D. Bruzewicz},  \textsc{R.~McConnell},
  \textsc{R.\,J. Ram},  \textsc{J.\,M. Sage},  and  \textsc{J.~Chiaverini},
 \jr{Nat. Nanotechnol.} \textbf{11}, 1066 (2016).


\bibitem{Langer05}
 \textsc{C.~Langer},  \textsc{R.~Ozeri},  \textsc{J.\,D. Jost},
  \textsc{J.~Chiaverini},  \textsc{B.~DeMarco},  \textsc{A.~Ben-Kish},
  \textsc{R.\,B. Blakestad},  \textsc{J.~Britton},  \textsc{D.\,B. Hume},
  \textsc{W.\,M. Itano},  \textsc{D.~Leibfried},  \textsc{R.~Reichle},
  \textsc{T.~Rosenband},  \textsc{T.~Schaetz},  \textsc{P.\,O. Schmidt},  and
  \textsc{D.\,J. Wineland},
 \jr{Phys. Rev. Lett.} \textbf{95}, 060502 (2005).


\bibitem{Brown2011}
 \textsc{K.\,R. Brown},  \textsc{A.\,C. Wilson},  \textsc{Y.~Colombe},
  \textsc{C.~Ospelkaus},  \textsc{A.\,M. Meier},  \textsc{E.~Knill},
  \textsc{D.~Leibfried},  and  \textsc{D.\,J. Wineland},
 \jr{Phys. Rev. A} \textbf{84}, 030303 (2011).


\bibitem{Gaebler2016}
 \textsc{J.\,P. Gaebler},  \textsc{T.\,R. Tan},  \textsc{Y.~Lin},
  \textsc{Y.~Wan},  \textsc{R.~Bowler},  \textsc{A.\,C. Keith},
  \textsc{S.~Glancy},  \textsc{K.~Coakley},  \textsc{E.~Knill},
  \textsc{D.~Leibfried},  and  \textsc{D.\,J. Wineland},
 \jr{Phys. Rev. Lett.} \textbf{117}, 060505 (2016).


\bibitem{Hsiang2014}
 \textsc{H.\,Y. Lo},  \textsc{J.~Alonso},  \textsc{D.~Kienzler},
  \textsc{B.\,C. Keitch},  \textsc{L.\,E. de~Clercq},  \textsc{V.~Negnevitsky},
   and  \textsc{J.\,P. Home},
 \jr{Appl. Phys. B} \textbf{114}, 17 (2014).


\bibitem{Brewer1988}
 \textsc{L.\,R. Brewer},  \textsc{J.\,D. Prestage},  \textsc{J.\,J. Bollinger},
   \textsc{W.\,M. Itano},  \textsc{D.\,J. Larson},  and  \textsc{D.\,J.
  Wineland},
 \jr{Phys. Rev. A} \textbf{38}, 859 (1988).


\bibitem{Wilson2011}
 \textsc{A.\,C. Wilson},  \textsc{C.~Ospelkaus},  \textsc{A.\,P. VanDevender},
  \textsc{J.\,A. Mlynek},  \textsc{K.\,R. Brown},  \textsc{D.~Leibfried},  and
  \textsc{D.\,J. Wineland},
 \jr{Appl. Phys. B} \textbf{105}, 741 (2011).


\bibitem{ball2013high}
 \textsc{H.~Ball},  \textsc{M.~Lee},  \textsc{S.~Gensemer},  and
  \textsc{M.~Biercuk},
 \jr{Rev. Sci. Instrum.} \textbf{84}, 063107 (2013).


\bibitem{cozijn2013laser}
 \textsc{F.~Cozijn},  \textsc{J.~Biesheuvel},  \textsc{A.~Flores},
  \textsc{W.~Ubachs},  \textsc{G.~Blume},  \textsc{A.~Wicht},
  \textsc{K.~Paschke},  \textsc{G.~Erbert},  and  \textsc{J.~Koelemeij},
 \jr{Opt. Lett.} \textbf{38}, 2370 (2013).


\bibitem{carollo2017third}
 \textsc{R.\,A. Carollo},  \textsc{D.\,A. Lane},  \textsc{E.\,K. Kleiner},
  \textsc{P.\,A. Kyaw},  \textsc{C.\,C. Teng},  \textsc{C.\,Y. Ou},
  \textsc{S.~Qiao},  and  \textsc{D.~Hanneke},
 \jr{Opt. Express} \textbf{25}, 7220 (2017).


\bibitem{Paschke2019}
 \textsc{A.\,G. Paschke},  \textsc{G.~Zarantonello},  \textsc{H.~Hahn},
  \textsc{T.~Lang},  \textsc{C.~Manzoni},  \textsc{M.~Marangoni},
  \textsc{G.~Cerullo},  \textsc{U.~Morgner},  and  \textsc{C.~Ospelkaus},
 \jr{Phys. Rev. Lett.} \textbf{122}, 123606 (2019).


\bibitem{Kuznetsov1997}
 \textsc{M.~Kuznetsov},  \textsc{F.~Hakimi},  \textsc{R.~Sprague},  and
  \textsc{A.~Mooradian},
 \jr{IEEE Photon. Technol. Lett.} \textbf{9}, 1063 (1997).


\bibitem{guina2017optically}
 \textsc{M.~Guina},  \textsc{A.~Rantam{\"a}ki},  and
  \textsc{A.~H{\"a}rk{\"o}nen},
 \jr{J. Phys. D} \textbf{50}, 383001 (2017).


\bibitem{myara2013}
 \textsc{M.~Myara},  \textsc{M.~Sellahi},  \textsc{A.~Laurain},
  \textsc{A.~Michon},  \textsc{I.~Sagnes},  and  \textsc{A.~Garnache},
 \jr{Proc. of SPIE} \textbf{8606}, 86060Q--1 (2013).


\bibitem{Burd:16}
 \textsc{S.\,C. Burd},  \textsc{D.\,T. Allcock},  \textsc{T.~Leinonen},
  \textsc{J.\,P. Penttinen},  \textsc{D.\,H. Slichter},  \textsc{R.~Srinivas},
  \textsc{A.\,C. Wilson},  \textsc{R.~J{\"o}rdens},  \textsc{M.~Guina},
  \textsc{D.~Leibfried},  and  \textsc{D.\,J. Wineland},
 \jr{Optica} \textbf{3}, 1294 (2016).


\bibitem{tropper2006}
 \textsc{A.~Tropper} and  \textsc{S.~Hoogland},
 \jr{Prog. Quantum. Electron.} \textbf{30}, 1 (2006).


\othercit
\bibitem{chilla2007}
 \textsc{J.~Chilla},  \textsc{Q.\,Z. Shu},  \textsc{H.~Zhou},
  \textsc{E.~Weiss},  \textsc{M.~Reed},  and  \textsc{L.~Spinelli},
Recent advances in optically pumped semiconductor lasers,
 in: Solid State Lasers XVI: Technology and Devices,  (2007),  p.\,645109.


\othercit
\bibitem{korpijarvi2015high}
 \textsc{K.~Ville-Markus},
PhD thesis, Tampere University of Technology, 2015.


\bibitem{Laurain2019}
 \textsc{A.~Laurain},  \textsc{J.~Hader},  and  \textsc{J.\,V. Moloney},
 \jr{J. Opt. Soc. Am. B} \textbf{36}(4), 847--854 (2019).


\bibitem{Castrillo2010}
 \textsc{A.~Castrillo},  \textsc{E.~Fasci},  \textsc{G.~Galzerano},
  \textsc{G.~Casa},  \textsc{P.~Laporta},  and  \textsc{L.~Gianfrani},
 \jr{Opt. Express} \textbf{18}, 21851 (2010).


\bibitem{Drever1983}
 \textsc{R.~Drever},  \textsc{J.~Hall},  \textsc{F.~Kowalski},
  \textsc{J.~Hough},  \textsc{G.~Ford},  \textsc{A.~Munley},  and
  \textsc{H.~Ward},
 \jr{Appl. Phys. B} \textbf{31}, 97 (1983).


\bibitem{le200575}
 \textsc{R.~Le~Targat},  \textsc{J.\,J. Zondy},  and  \textsc{P.~Lemonde},
 \jr{Opt. Commun.} \textbf{247}, 471 (2005).


\bibitem{kretzschmar20172}
 \textsc{N.~Kretzschmar},  \textsc{U.~Eismann},  \textsc{F.~Sievers},
  \textsc{F.~Chevy},  and  \textsc{C.~Salomon},
 \jr{Opt. Express} \textbf{25}, 14840 (2017).


\bibitem{Cook2018}
 \textsc{E.\,C. Cook},  \textsc{A.\,D. Vira},  \textsc{C.~Patterson},
  \textsc{E.~Livernois},  and  \textsc{W.\,D. Williams},
 \jr{Phys. Rev. Lett.} \textbf{121}, 053001 (2018).


\othercit
\bibitem{Demtroder}
 \textsc{W.~Demtr{\"o}der},
Spectroscopic Instrumentation (Springer, 1981).


\othercit
\bibitem{Tan2014}
 \textsc{T.\,R. Tan},
PhD thesis, University of Colorado at Boulder, 2016.


\bibitem{kantola2014}
 \textsc{E.~Kantola},  \textsc{T.~Leinonen},  \textsc{S.~Ranta},
  \textsc{M.~Tavast},  and  \textsc{M.~Guina},
 \jr{Opt. Express} \textbf{22}, 6372 (2014).


\end{thebibliography}
	
\end{document}